# An Overview of the 3GPP Study on Artificial Intelligence for 5G New Radio

Xingqin Lin

NVIDIA

Email: xingqinl@nvidia.com

*Abstract*— Air interface is a fundamental component within any wireless communication system. In Release 18, the 3rd Generation Partnership Project (3GPP) delves into the possibilities of leveraging artificial intelligence (AI)/machine learning (ML) to improve the performance of the fifth-generation (5G) New Radio (NR) air interface. This endeavor marks a pioneering stride within 3GPP's journey in shaping wireless communication standards. This article offers a comprehensive overview of the pivotal themes explored by 3GPP in this domain. Encompassing a general framework for AI/ML and specific use cases such as channel state information feedback, beam management, and positioning, it provides a holistic perspective. Moreover, we highlight the potential trajectory of AI/ML for the NR air interface in 3GPP Release 19, a pathway that paves the journey towards the sixth generation (6G) wireless communication systems that will feature integrated AI and communication as a key usage scenario.

## I. INTRODUCTION

The 3rd Generation Partnership Project (3GPP) has successfully concluded the initial phase of fifth-generation (5G) advancement through its Releases 15–17. Now, it embarks on the subsequent stage of 5G evolution, referred to as 5G-Advanced [1]. As the first release of 5G-Advanced, Release 18 includes comprehensive projects which not only cater to immediate commercial needs but also encompass long-term endeavors that lay the groundwork for the evolution of wireless access into the realms of the sixth generation (6G) [2]. A notable focal point within this ambit is the integration of artificial intelligence (AI)/machine learning (ML) into the fabric of 5G-Advanced evolution, which is set to facilitate the widespread adoption of AI/ML within wireless communication systems.

Prior to the advent of 5G-Advanced evolution, 3GPP engaged in preliminary AI/ML initiatives during the first phase of 5G evolution, spanning multiple domains from the 5G core network (5GC), the operations, administration, and maintenance (OAM), and the radio access network (RAN) [3][4]. Air interface stands as a fundamental component within any wireless communication system. Recent works have painted visions of an AI-native air interface (see, e.g., [5][6]). Nevertheless, it's noteworthy that the preliminary AI/ML endeavors conducted by 3GPP prior to the 5G-Advanced evolution did not cover the 5G New Radio (NR) air interface. 3GPP addresses this gap in Release 18 by exploring the potential of using AI/ML-based algorithms to enhance the NR air interface [7]. This pursuit marks a groundbreaking stride as it is the first of its kind within 3GPP's development of wireless communication standards. The scope of this exploration includes the development of a general AI/ML framework, alongside the exploration of specific use cases such as channel state information (CSI) feedback, beam management, and positioning [8].

The surge of interest in employing AI/ML within wireless communication systems has catalyzed the publication of a number of papers [9]-[15]. Overviews of recent advances and future challenges for AI/ML-enabled wireless networks are provided in [9][10]. But these works only offer brief discussions on the standardization efforts at the time of 2020, with a notable absence of coverage regarding the latest strides made in 3GPP Release 18 which commenced in 2022. The potential of implementing AI/ML solutions within the open RAN (O-RAN) architecture is explored in [11], but the study confines its scope to the O-RAN Alliance, leaving a gap concerning the ongoing 3GPP standardization endeavors. A more recent work [12] introduces the 3GPP Release-18 study by solely focusing on the use case of using AI/ML for CSI feedback. The other two use cases, i.e., AI/ML-based beam management and AI/ML-based positioning, are investigated in [13] and [14], respectively. Nonetheless, despite these use case-specific investigations, a comprehensive introduction to the 3GPP Release-18 study on AI/ML for the NR air interface is still lacking. Our recent work [15] includes a section which presents a brief overview of the 3GPP Release-18 study, but it does not delve into the detailed progresses made by 3GPP.

In this article, we address a critical gap within the existing literature by offering a dedicated treatment of the 3GPP Release-18 study on AI/ML for the NR air interface. This comprehensive contribution enriches the existing works which often confine their scopes to certain areas of the 3GPP study. Furthermore, we also divulge the myriad factors that underpin the process of standardization. Given that the integration of AI/ML within the air interface is a nascent and largely uncharted avenue in the realm of standards development, 3GPP has identified many new challenges and gained novel perspectives during this study. Conveying these learnings from the front lines of 3GPP helps demystify the decisions and is particularly valuable for researchers not directly involved in the 3GPP work.

## II. GENERAL AI/ML FRAMEWORK

To establish a general 3GPP AI/ML framework for the air interface, considerable efforts from 3GPP members have been dedicated to formulating shared terminology that pertains to AI/ML functions, procedures, and interfaces. Figure 1 outlines the 3GPP AI/ML functional framework for the NR air interface. The framework delineates a set of core functions, including data collection, model training, management, inference, and model storage.

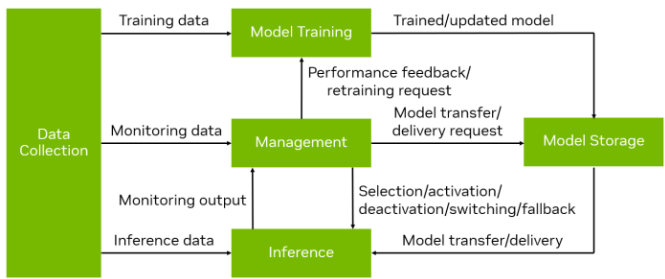

**Figure 1: 3GPP AI/ML functional framework for the NR air interface.**

An AI/ML model needs to be developed, deployed, and managed during the entire lifecycle—a process known as AI/ML model life cycle management (LCM). 3GPP has studied two distinct methods for managing the life cycle of an AI/ML model at user equipment (UE). The first method is categorized as *functionality-based LCM*. A functionality refers to an AI/ML-enabled feature or feature group facilitated by a configuration. AI/ML functionality identification fosters mutual understanding between the network and the UE about the AI/ML functionality. The process of functionality identification may be integrated within the existing UE capability signaling framework. Essentially, configurations are tailored in accordance with conditions indicated by UE capability. Subsequently, upon identifying functionalities, the UE can report updates pertaining to the applicable functionalities among those configured or identified. In functionality-based LCM, the network indicates selection, activation, deactivation, switch, and fallback of an AI/ML functionality through 3GPP signaling. Notably, the exact AI/ML model(s) that underpin a given functionality might not be identified at the network.

The second method is categorized as *model identity (ID) based-LCM*. A model ID serves as a distinctive identifier for an AI/ML model, wherein the model could be logical and its mapping to a physical model is up to implementation. AI/ML model identification ensures a mutual understanding between the network and the UE concerning the AI/ML model in question. Specifically, the AI/ML model is identified by its designated model ID at the network, and the UE indicates its supported AI/ML model to the network. Besides the model ID, the model can have accompanying conditions as part of the UE capability definition as well as additional conditions (e.g., scenarios, sites, and datasets) which determine the applicability of the model. In model-ID-based LCM, both the network and the UE may perform selection, activation, deactivation, switch, and fallback of an AI/ML model by using the corresponding model ID.

The commercial deployment of an AI/ML-enabled feature hinges on its ability to deliver reliable performance across a spectrum of scenarios, configurations, and site-specific conditions in mobile communication systems. To achieve this objective, 3GPP has investigated three approaches: model generalization, model switching, and model update. Model generalization aims to develop one model generalizable to different scenarios, configurations, or sites. Alternatively, a set of specific models can be developed—ranging from scenario-specific to configuration- or site-specific. Within this spectrum of models, the technique of model switching is harnessed to effectively address the different scenarios, configurations, or sites. The process of model update, often involving fine-tuning, entails a flexible adaptation of the model structure or its parameters in response to changes in scenarios, configurations, or sites. A pivotal principle underpinning these approaches is to ensure that the performance of AI/ML-enabled features remains at a level equal to or better than that of legacy non-AI/ML-based operations. Therefore, performance monitoring is a must for the AI/ML-enabled features, calling for functions such as computing monitored performance metrics, reporting monitoring results, and control signaling mechanisms to swiftly recover from failure.

Different use cases require varying degrees of collaboration between the network and the UE for the corresponding AI/ML operations. 3GPP has identified three distinctive levels of network-UE collaboration:

- **Level x–no collaboration:** At this level, AI/ML operations are grounded in proprietary implementations devoid of any specific standards enhancements tailored for AI/ML functionalities.
- **Level y–signaling-based collaboration without model transfer:** At this level, AI/ML operations integrate dedicated standards enhancements to facilitate the process without involving model transfer. Here, 'model transfer' refers to the delivery of an AI/ML model over the air interface from one entity to another, conducted in a manner not transparent to 3GPP signaling mechanisms.
- **Level z–signaling-based collaboration with model transfer:** Within this tier, the AI/ML operations encompass not only the integration of new signaling but also leverage advanced model transfer capabilities.

In essence, these collaboration levels encompass a spectrum from minimal involvement to deep integration, signifying the versatility and adaptability of AI/ML operations across different contexts.

## III. USE CASE: CSI FEEDBACK

CSI refers to the information of the multipath wireless channel between a 5G node B (gNB) and a UE. The UE can measure downlink reference signals, compute downlink CSI, and provide a CSI report to the gNB, thereby facilitating downlink transmission. Leveraging AI/ML-based algorithms can further enhance CSI feedback, yielding advantages such as reduced overhead and improved accuracy. 3GPP has studied two representative sub-use cases of employing AI/ML-based algorithms for CSI feedback enhancement: spatial-frequency domain CSI compression and time domain CSI prediction with a UE-sided model.

### A. Spatial-frequency Domain CSI Compression

In AI/ML-based CSI compression, a UE employs an AI/ML-based CSI encoder to generate CSI feedback information, while a corresponding AI/ML-based CSI decoder at the gNB is used to reconstruct the CSI from the received feedback data. This is an example of a *two-sided AI/ML model*, where the inference operation is split between the UE and the gNB. This two-sided AI/ML model can be employed to compress either the raw channel matrix estimated by the UE or the precoding matrix derived from the raw channel matrix. Notably, compressing the precoding matrix aligns with the existing codebook-based CSI feedback framework specified for the NR air interface, thus attracting more interest during the 3GPP study.

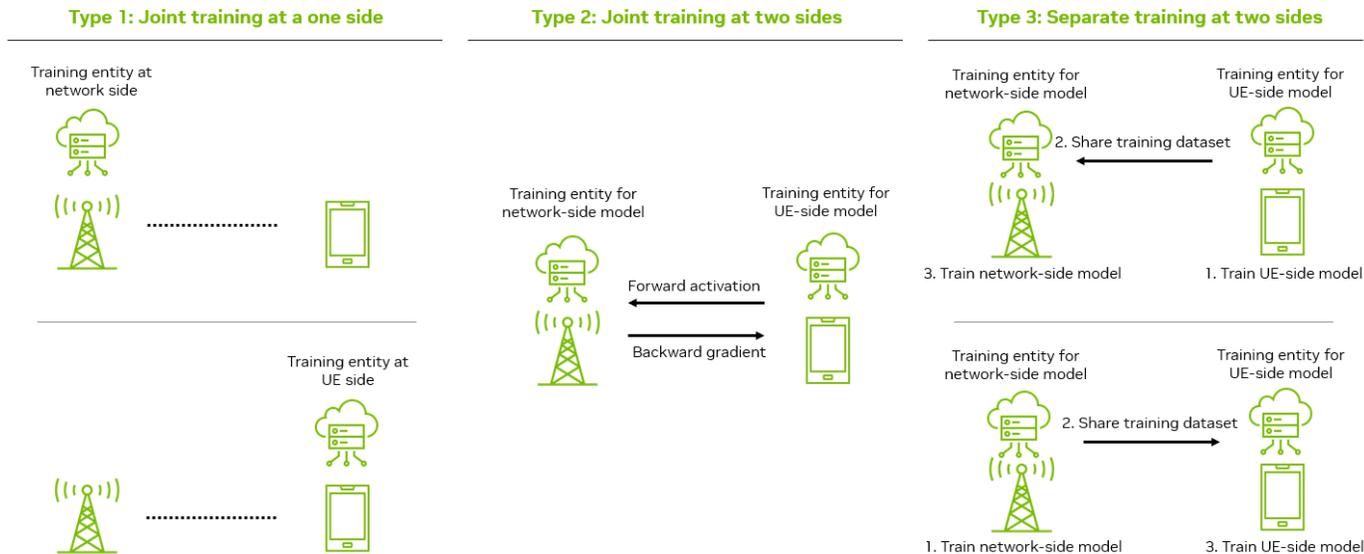

Figure 2: Types of AI/ML model training for CSI compression using a two-sided model.

Using a two-sided AI/ML model for the air interface introduces a multitude of challenges. The first challenge involves the training of the two-sided AI/ML model. Within this context, 3GPP has investigated three types of training that involve varying degrees of collaboration between the network and the UE, as illustrated in Fig. 2. In the first training scenario, designated as 'Type 1,' the encoder and decoder of the two-sided model are jointly trained at one side. If this training is done at the network side, the resultant encoder model can be transferred to the UE, and vice versa. Moving on to 'Type 2,' the encoder and decoder of the two-sided model undergo training at the UE side and network side, respectively. Precisely, the encoder and decoder are jointly trained in the same loop for forward propagation and backward propagation by exchanging forward activation and backward gradient between the UE and the network. Lastly, 'Type 3' encompasses the separate training of the encoder and decoder within the two-sided model, conducted in distinct training sessions at the UE and network sides. The separate training can start at the UE side or the network side. Take the separate training starting at the UE side for example. In the initial phase, a UE-side entity trains the CSI encoder model. Subsequently, the UE-side entity shares a training dataset—comprising encoder outputs and target CSI—with a network-side entity. This dataset is then leveraged by the network-side entity to train its CSI decoder model.

Different training approaches offer different advantages and challenges. Among these, training type 1 holds the promise of yielding superior performance, but the deployment and management of the models can be intricate. Take training type 1 at the network side for example. The network side entity shall involve the UE vendor of the targeted UE type in the training so that the trained CSI encoder model is compatible with the UE's implementation. However, this effort entails the disclosure of proprietary information by the UE vendor to the network side, which can be a significant challenge. On the other hand, training type 2 retains the confidentiality of proprietary information at both the network and UE sides, as model information remains within their respective domains. Nonetheless, the two respective training entities need to coordinate their training iterations to exchange forward activation and backward propagation results, which can lead to significant coordination effort and overhead. Training type 3 emerges as an option that not only safeguards proprietary information but also eliminates the necessity for collaboration during the training iterations because coordination can take place outside the training process.

The training complexity inherent in a two-sided AI/ML model for the air interface is further compounded by the necessity for multi-vendor interoperability and compatibility. The CSI decoder located at the gNB needs to be compatible with different CSI encoders at the UEs, and vice versa. In scenarios where a common CSI decoder model is utilized for multiple CSI encoder models, the network-side training entity—under training type 1 or 2—must coordinate with UE vendors for joint training efforts. Notably, the release of a new UE type could potentially trigger retraining across all vendors. Similar challenges exist in training type 1 or 2 when a shared CSI encoder model is used for multiple CSI decoder models. These issues can be mitigated in training type 3. In particular, if a common CSI decoder model is used for multiple CSI encoder models (or vice versa), the retraining due to the release of a new UE type can be confined to involve only the associated UE vendor and the network vendor due to the separate training nature in training type 3. In summary, for CSI feedback hinging on a two-sided AI/ML model, training type 3 emerges as a more pragmatic and feasible approach in comparison to training types 1 and 2.

Another key consideration in the design of AI/ML-based CSI compression is the generalization capability of the AI/ML model across a spectrum of scenarios and configurations. Specifically, the design needs to consider diverse scenarios, including different deployment scenarios (e.g., urban macro, urban micro, and indoor), UE distributions, carrier frequencies, etc. The AI/ML model should also be scalable over various configurations, such as different channel bandwidths, frequency granularities, CSI feedback payloads, as well as variations in antenna port layouts and numbers. These diverse configurations can impact the AI/ML model design as they may lead to different dimensions of model input and output.

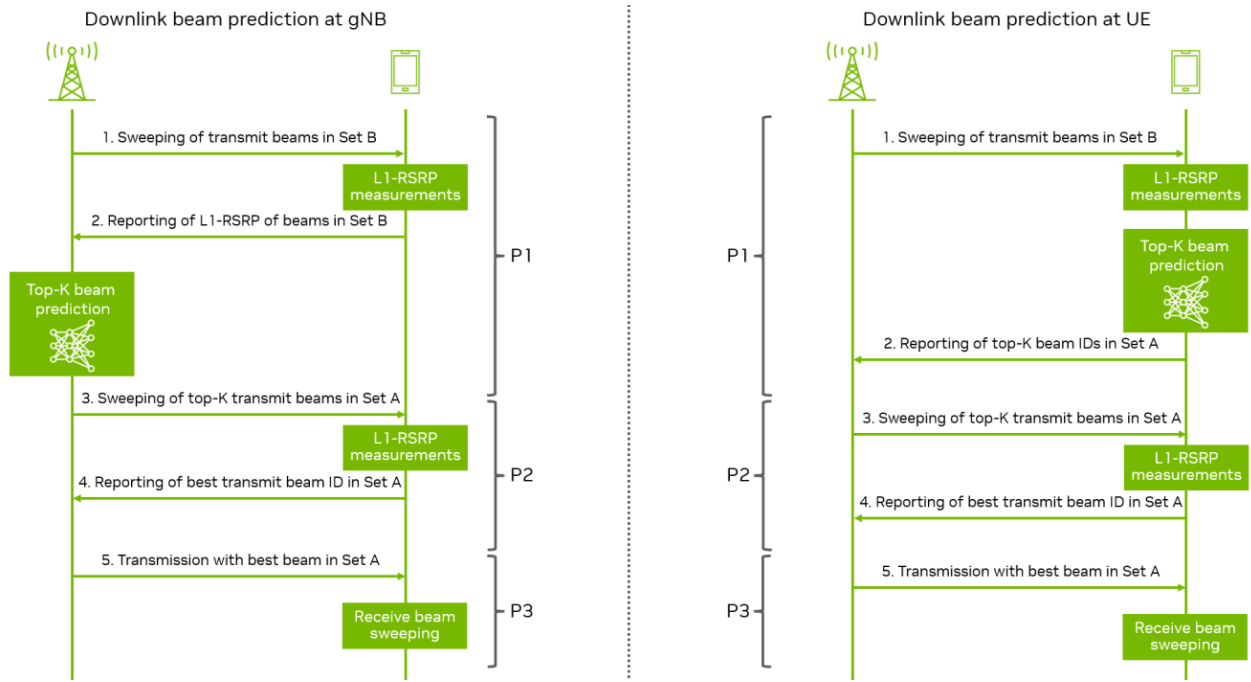

**Figure 3: Downlink beam management procedures with AI/ML-based beam prediction at gNB side (left) and at UE side (right).**

*B. Time Domain CSI Prediction with a UE-sided Model*

A challenge within the legacy CSI reporting framework of NR pertains to a temporal lag between the time to which the reported CSI corresponds and the moment at which the gNB actually employs the CSI report. This time delay leads to a situation where the reported CSI becomes outdated, a phenomenon commonly referred to as channel aging. Notably, the pace at which the reported CSI becomes outdated is amplified by higher UE speeds. This concern becomes particularly pronounced in the context of multi-user multiple-input multiple-output (MU-MIMO) scenarios, especially within massive MIMO deployments. The performance of MU-MIMO has been observed to deteriorate when UEs move at medium to high speeds. Leveraging AI/ML algorithms for CSI prediction emerges as a promising technique to counteract the effect of outdated CSI.

In contrast to AI/ML-based CSI compression, which necessitates a two-sided model, AI/ML-based CSI prediction in the time domain can employ a one-sided model. Training this one-sided model can be executed by a single vendor, and the inference can subsequently be conducted by one side (either the gNB or the UE). Considering the workload in Release 18, 3GPP has strategically focused on the UE-sided model for CSI prediction in the time domain. In this setup, the input for the one-sided model consists of a sequence of past CSI measurements taken by the UE. The resultant output of this model is a forecasted CSI for a future time instance, predicted by the UE.

From the standpoint of 3GPP standards, it is expected that we can largely reuse the existing CSI framework to support CSI prediction. In particular, the AI/ML model LCM for UE-sided CSI prediction can to a large extent reuse what is defined for other UE-sided use cases since the specification impact of UE-sided AI/ML models have already been investigated for beam management and positioning, as described in the following sections.

## IV. USE CASE: BEAM MANAGEMENT

Beam management functionality in NR is used to support beamforming. It is particularly needed for 5G millimeter wave systems that rely on analog beamforming. In a basic procedure for downlink beam management, the UE measures the reference signal associated with each gNB transmit beam and tests different UE receive beams for each gNB transmit beam to find a suitable downlink beam pair. This process can be time-consuming and entail a substantial overhead in terms of reference signals. AI/ML-based algorithms offer the potential to enhance beam management functionality, delivering advantages that include the reduction of overhead, minimized latency, and improved accuracy in beam selection.

3GPP has studied two representative sub-use cases that involve the application of AI/ML-based algorithms to beam management. They are referred to as 'spatial-domain downlink beam prediction' and 'time-domain downlink beam prediction.'

- **Spatial-domain downlink beam prediction** leverages measurement outcomes from a designated set of downlink beams, denoted as 'Set B,' to predict the best beam within another set of downlink beams, termed 'Set A,' at the present moment.
- **Time-domain downlink beam prediction** harnesses historical measurement results derived from 'Set B' to anticipate the best beam in 'Set A' for one or more future time instances.

Notably, 'Set B' can either constitute a subset of 'Set A,' or the two sets can be different (e.g., 'Set B' comprising wide beams while 'Set A' is composed of narrow beams). Additionally, in the context of time-domain downlink beam prediction, 'Set A' and 'Set B' can also be the same.

A typical input to an AI/ML model for the spatial- or time-domain downlink beam prediction is layer 1 reference signal received power (L1-RSRP) measurements of beams within 'Set B.' A typical output from the AI/ML model is the predicted top-K beams in 'Set A.' The AI/ML model training and inference can reside at the gNB side or at the UE side. In scenarios where AI/ML inference occurs at the UE side, the UE needs to report its predicted beam(s) to the gNB. Alternatively, when AI/ML inference takes place at the gNB side, the UE is required to report its L1-RSRP measurements for the beams within 'Set B' to the gNB.

To provide clarity regarding AI/ML-based downlink beam prediction, we now elaborate on how beam prediction can be integrated into the existing NR beam management framework. Recall that the existing framework consists of three procedures known as P1 (initial beam pair establishment), P2 (transmit beam refinement), and P3 (receive beam refinement). Figure 3 provides an illustration of downlink beam management procedures with AI/ML-based beam prediction at the gNB side and at the UE side.

Take beam prediction at the gNB side for example. The gNB first sweeps through different transmit beams in 'Set B,' which are measured by the UE. Subsequently, the UE conveys the L1-RSRP measurements for beams within 'Set B' back to the gNB. Upon reception of the report, the gNB uses the received L1-RSRP values as input to its AI/ML model, which then makes prediction about the top-K beams within 'Set A.' Based on the inference result, the gNB then sweeps through the predicted top-K transmit beams within 'Set A,' which the UE measures to identify the best transmit beam that yields the highest L1-RSRP value. Following this, the UE reports the ID of the best beam back to the gNB. Should the need arise, the gNB has the option to trigger the P3 procedure. In this phase, the gNB keeps using the best transmit beam, while the UE probes different receive beams to determine the most suitable one.

## V. USE CASE: POSITIONING

The existing 5G NR positioning methods are typically geometry-based, consisting of two main steps: 1) conducting measurements of radio signals, and 2) calculating a position estimate by solving a system of non-linear equations that establish a relationship between the UE's position and the measurements. The accuracy of the geometry-based positioning methods heavily depends on the availability of measurements linked with line-of-sight (LOS) paths. In scenarios involving weak LOS conditions or dense multipath environments, such as indoor factory settings, the accuracy of geometry-based methods tends to degrade. AI/ML-based algorithms can enhance positioning accuracy across a range of scenarios, including those characterized by prevalent non-LOS (NLOS) conditions.

3GPP has studied two representative sub-use cases involving the application of AI/ML-based algorithms to positioning. These have been termed as 'direct AI/ML positioning' and 'AI/ML-assisted positioning.'

- **Direct AI/ML positioning** employs an AI/ML model to directly determine the location of UE. For instance, this can encompass fingerprinting-based positioning utilizing channel observations, such as channel impulse response (CIR) or power delay profile (PDP), as input to the AI/ML model.
- **AI/ML-assisted positioning** involves leveraging an AI/ML model to generate an intermediate measurement statistic, which is instrumental in positioning. This could encompass measurements such as LOS/NLOS probability, angle-of-arrival/departure, or time-of-arrival.

The AI/ML model training and inference can reside at the UE side, the location management function (LMF) side, or the gNB side. Depending on the roles of UE, LMF, and gNB in the positioning procedures, 3GPP focused on three categories of positioning methods. The first category is UE-based positioning, where the UE itself executes either direct AI/ML positioning or AI/ML-assisted positioning. The second category is UE-assisted LMF-based positioning, where the UE provides assistance to the LMF in estimating the UE's location. In this scenario, the LMF can perform direct AI/ML positioning, or the UE can engage in AI/ML-assisted positioning. The third category is gNB-assisted positioning, where the gNB provides assistance to the LMF in estimating the UE's location. In this case, the LMF can implement direct AI/ML positioning, or the gNB can participate in AI/ML-assisted positioning.

Exploring the generalization capability of AI/ML models is a key area of investigation across the use cases in 3GPP Release 18. As far as AI/ML-based positioning is concerned, various dimensions have been considered to investigate the model generalization capability. For example, training and test datasets are generated under different drops, varying clutter parameters, or distinct timing errors. To help make sense of this issue, the left part of Fig. 4 presents a case study illustrating the generalization capability of direct AI/ML positioning, using the 3GPP simulation setup [7]. The simulated deployment scenario is indoor factory with heavy NLOS conditions. The simulated AI/ML model takes the form of a convolutional neural network (CNN), where CIRs serve as model input and predicted UE position constitutes the model output. Two distinct datasets, denoted as 'drop 1' and 'drop 2,' were generated using different random seed values and used for model training and testing, respectively. In essence, these two drops can be envisioned as representing two different indoor factories that have the same clutter settings. The results reveal a notable decline in the precision of direct AI/ML positioning when training and testing are executed on different drops. One plausible remedy for this problem is to use AI/ML model fine-tuning, as also shown in the left part of Fig. 4. It is worth noting that the original model was initially trained with a dataset of 16,000 samples. Therefore, the employment of 1,000 (or 2,000) fine-tuning samples translates to 6.25% (or 12.5%) of the 16,000 samples needed for training the AI/ML model from scratch.

Data collection is another key area of exploration across the use cases studied in 3GPP Release 18. When considering AI/ML-based positioning, there is an intuitive expectation that an increased density of gathered data would lead to enhanced positioning accuracy. This intuition is quantitatively illustrated in the right part of Fig. 4. The discernible trend reveals that the improvement of positioning accuracy comes at the cost of high requirement on data collection. This underscores the importance of data collection strategies for the effective implementation of AI/ML-based positioning.

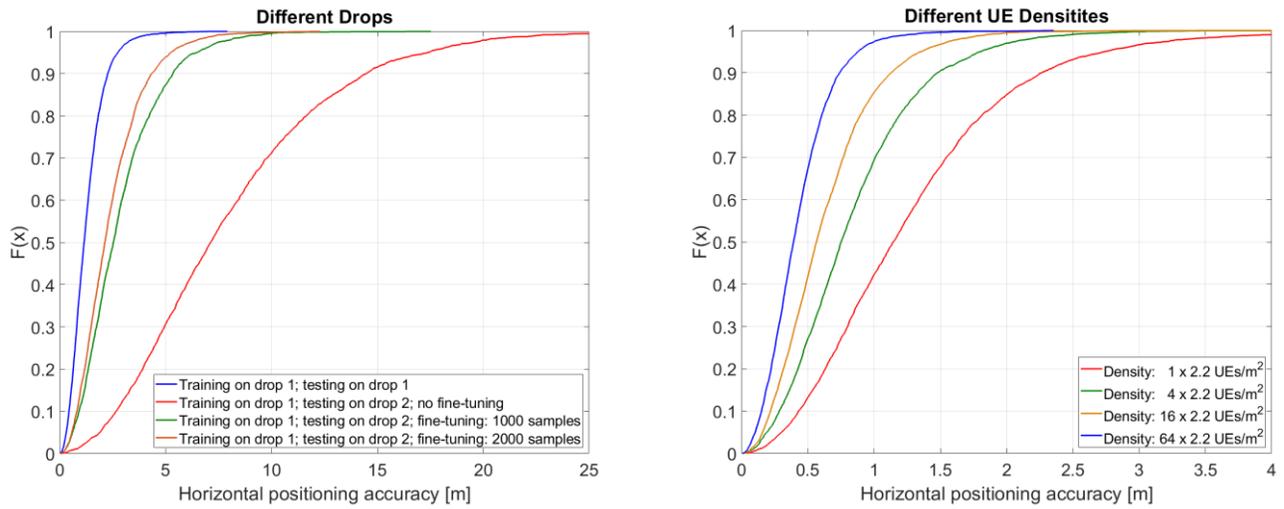

**Figure 4: Positioning accuracy of direct AI/ML positioning under different drops (left) and different UE densities (right).**

## VI. INTEROPERABILITY AND TESTABILITY

Interoperability and testability are critical considerations in the development and deployment of standardized features within cellular networks, including AI/ML-based schemes. 3GPP RAN working group 4 (RAN4), responsible for setting performance requirements and defining test procedures, is investigating the interoperability and testability aspects for validating AI/ML-based performance enhancements. The incorporation of AI/ML into the air interface introduces significant challenges to the existing requirements and testing framework. Notably, AI/ML models are data-driven and often lack physical interpretations, rendering the prediction of their performance difficult. In this section, we scratch the surface of this largely uncharted territory by highlighting the key areas under development in 3GPP.

The scope of 3GPP RAN4 requirements and testing for AI/ML-based features encompasses a range of vital elements, including inference, LCM procedures, data generation and collection, and generalization verification. Core requirements include the performance monitoring procedure, functionality/model management procedure, and the corresponding latency and interruption requirements. 3GPP considers a reference block diagram for testing AI/ML-based features, as illustrated in Fig. 5. Within this framework, the device under test (DUT) can be either UE or gNB. The reference block diagram covers both one-sided and two-sided models. In the latter case, the test equipment incorporates a companion AI/ML model to perform joint inference with the model at the DUT. However, the methodology for devising a reference AI/ML model within the testing equipment to effectively test the performance of the corresponding AI/ML model within the DUT remains an ongoing topic of discussion.

While the two-sided models pose more interoperability challenges, it is important to note that interoperability considerations also extend to the utilization of one-sided AI/ML models. This includes aspects such as procedure signaling and testing setups to ensure compliance with minimum requirements. Traditionally, 3GPP RAN4 defines requirements for testing equipment in controlled laboratory conditions prior to its field deployment. Generalization verification is an intricate task. In particular, the training of an AI/ML model can be tailored to encompass the entirety of conditions outlined in the standard, thus demonstrating superior performance during testing. However, the AI/ML model may be overfitted to the standardized setups, possibly compromising its robustness in real-world environments that are not aligned with the controlled conditions. Moreover, AI/ML models might need periodic updates even after they are deployed in live networks. These new considerations call for the implementation of performance monitoring mechanisms to detect non-compliance, as well as the formulation of new testing procedures to effectively validate the functionality of AI/ML-based features operating in the field.

## VII. CONCLUSION AND FUTURE OUTLOOK

The 3GPP Release-18 study on AI/ML for the NR air interface is a pioneering initiative in the 3GPP's development of wireless communication standards. This article has timely offered an overview of the key topics investigated by 3GPP in this area. Considering that this is a largely uncharted territory for standards development, substantial work remains ahead within 3GPP to cultivate this domain into a state of maturity fit for commercial deployments at scale. 3GPP has already initiated discussions pertaining to the subsequent release, namely Release 19. The support within the ecosystem for advancing AI/ML initiatives for the air interface in Release 19 is strong. It is envisaged that 3GPP will conduct normative work on AI/ML for the NR air interface based on the outcomes of the Release-18 study. Additionally, new use cases such as AI/ML-based mobility management will be explored, and further studies will delve into areas that warrant deeper investigation, such as testing methodologies for two-sided

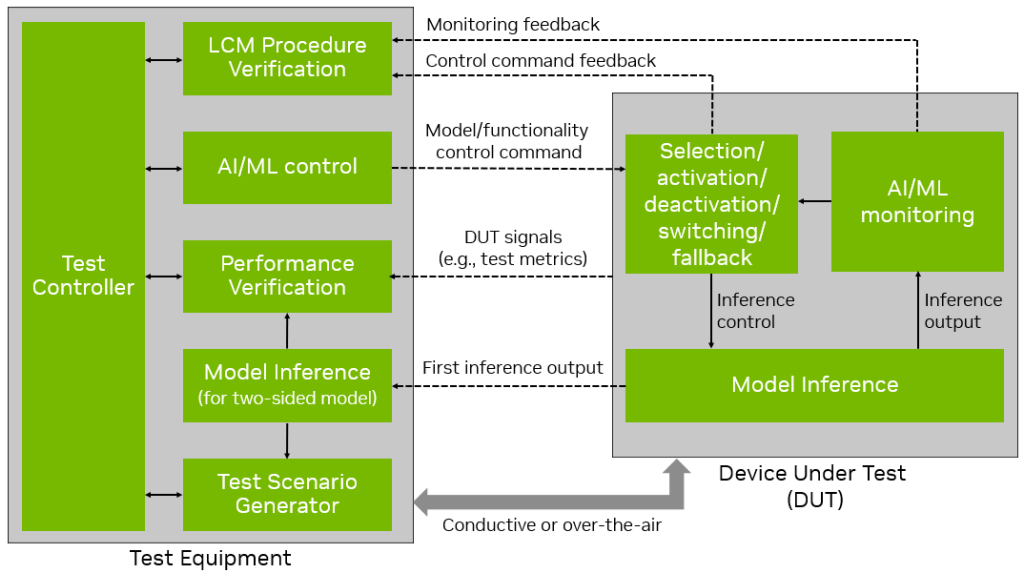

**Figure 5: Reference block diagram for testing AI/ML-based features.**

AI/ML models. These concerted efforts are poised to lay the foundation for the forthcoming 6G that will feature integrated AI and communication as a key usage scenario. As the AI and wireless communication landscapes coalesce, the groundwork laid by 3GPP will undoubtedly shape the contours of this transformative frontier.